\documentclass[10pt,aps,pre,twocolumn,amsmath,amsfonts,amssymb,floatfix,superscriptaddress,notitlepage,nofootinbib]{revtex4-1}
\usepackage{amssymb,amsmath}
\usepackage{bm}
\usepackage{color}
\usepackage{hyperref}
\usepackage{soul}

\usepackage{amsmath}
\usepackage{amsfonts}
\usepackage{graphicx}
\usepackage{color}
\usepackage{indentfirst} 
\usepackage{enumitem}

\begin{document}
\title{Computational Crystallization}
\author{Irem Altan} 
\affiliation{Department of Chemistry, Duke University, Durham, NC 27708, USA}
\author{Patrick Charbonneau}
\affiliation{Department of Chemistry, Duke University, Durham, NC 27708, USA}
\affiliation{Department of Physics, Duke University, Durham, NC 27708, USA}
\author{Edward H. Snell}
\affiliation{Hauptman-Woodward Medical Research Institute, 700 Ellicott St. NY 14203, USA}
\affiliation{Department of Structural Biology, SUNY University of Buffalo, 700 Ellicott St. NY 14203, USA}

\begin{abstract}
Crystallization is a key step in macromolecular structure determination by crystallography. While a robust theoretical treatment of the process is available, due to the complexity of the system, the experimental process is still largely one of trial and error. In this article, efforts in the field are discussed together with a theoretical underpinning using a solubility phase diagram. Prior knowledge has been used to develop tools that computationally predict the crystallization outcome and define mutational approaches that enhance the likelihood of crystallization. For the most part these tools are based on binary outcomes (crystal or no crystal), and the full information contained in an assembly of crystallization screening experiments is lost. The potential of this additional information is illustrated by examples where new biological knowledge can be obtained and where a target can be sub-categorized to predict which class of reagents provides the crystallization driving force. Computational analysis of crystallization requires complete and correctly formatted data. While massive crystallization screening efforts are under way, the data available from many of these studies are sparse. The potential for this data and the steps needed to realize this potential are discussed.
\end{abstract}

\maketitle

\section{Introduction}
Macromolecular crystallography is a gateway to the detailed structure of biological molecules, and, to the biological processes in which they are involved. Its power as a scientific tool is recognized in the remarkable number of Nobel prizes that make direct reference to the technique. The main drawback of the approach is that – as its name indicates – it requires a crystal of the macromolecule of interest. There lies the crux of the problem.

The National Institutes of Health Protein Structure Initiative (PSI) targeting fold space (where all outcomes, crystallization and non-crystallization, were tracked) showed that out of $\sim$45K soluble, purified targets, $\sim$8K crystallized, of which only $\sim$5K resulted in a crystal structure. Later data from the PSI-Biology initiative, which focused on targets of compelling biological interest, showed that out of a further $\sim$10K targets, only $\sim$2K resulted in a crystal structure \cite{w1}. Results from these large datasets mirror those obtained at the large-scale crystallization screening center led by one of us: the Hauptman-Woodward Medical Research Institute High-Throughput Screening (HTS) Laboratory \cite{w2}.  An analysis of 96 biological macromolecular targets screened against a set of 1536 chemical cocktails gave 277 crystal leads (36 targets produced one or more crystals) from $\sim$150K experiments, and hence only $\sim$0.2\% of the experiments used for screening produced crystals \cite{w3,w4}. These results thus provide a consistent illustration of the remarkably poor success rate of crystallization screening despite the power of the crystallographic technique.

A meta-analysis of large-scale crystallization screening centers reveals that out of ten soluble protein constructs, four are likely to crystallize, of which only one on average will produce a crystal structure \cite{w5}. Because the number of structures deposited in the Protein Data bank (PDB) that have been determined by X-ray crystallographic methods is now over 100K, we infer that tens of millions of experiments have had outcomes other than diffracting crystals. Unfortunately, information on those experiments is not reported in the PDB, and in most cases nor are the conditions that produced a crystal but were not used in the final structure determination. We are thus left with datasets that incompletely capture the effort involved in the crystallization of macromolecules. If that were not bad enough, the data that is available often presents multiple difficulties for automated attempts at data mining and prediction. In this article, we first describe the crystallization process, and then show that by utilizing the PDB-deposited structures and sequences along with the information whether or not that sequence produced a crystal, it is possible to provide predictive glimpses of the experimental outcome. We then review attempts at employing detailed screening information in order to provide further insight. We conclude with a discussion of the incentives for improved bookkeeping and with an outlook towards future research in computational crystallization.

\section{Crystallization}
\label{sec:crys}
Crystallization has been described as an ``empirical art of rational trial and error guided by past results''\cite{w6}. For better and for worse, this still sums up the state of the art. When presented with a biological macromolecular sequence it is currently impossible to know what conditions will result in crystallization or even if crystallization will occur at all. Basic predictive techniques can determine, for instance, if the target may be associated with a membrane, providing an initial lead into the appropriate crystallization approaches, but little further guidance is available. The history of crystallization methods and the development of modern crystallization screens have been summarized elsewhere \cite{w7}. It is, however, important to note that the development of readily prepared commercial screens make initial crystallization screening experiments easier, but at the risk of standardizing the starting points and introducing a bias on outcome. We will not discuss the details of the available screens or approaches here, but rather sketch the features of these methods that are essential to describing the role and purpose of experimental crystallization data. 

The main strategy in macromolecular crystallization is to gradually bring a target solubilized in an appropriate aqueous chemical cocktail (containing at least a precipitating agent and a buffer) to a region of supersaturation until a crystal nucleates and grows.  Interestingly, experimental outcomes reveal that there are many landmarks on the screening landscape that are not crystals. Crystals form under solution conditions that fall between those producing clear drops (conditions under which macromolecular-solvent interactions are stronger than interactions between macromolecules) and those producing precipitate (conditions under which the opposite is true). Clear and precipitate are common outcomes but phase separation, skin-formation, and a combination of any of these outcomes can also occur. These outcomes are markers on the solubility landscape that can be used to direct experiments towards a crystal \cite{w8}. For the vast majority of reports, however, these outcomes are not captured, restricting significantly the insights that can be gained from failed crystallization experiments into profiling the phase response of a macromolecule to different chemistries.

To understand the potential insights it is useful to review the thermodynamical framework of crystallization screening experiments. First and foremost, macromolecular crystallization is a phase transition. To study the coexistence of the crystal and the solution forms of a macromolecule, one needs to determine the set of conditions at which the chemical potentials of the two phases are equal. The chemical potential determination from first-principles entails knowledge of effective interactions between macromolecules \cite{w9,w10,w11,w12}. Unfortunately, although the individual forces involved in macromolecular interactions are well understood, there does not exist a simple and fast way to figure out how these forces collectively build up effective macromolecular interactions \cite{w13,w14}.

Both experimental and theoretical studies of crystallization inevitably involve the interpretation of phase diagrams. These recapitulate the conditions under which different phases are thermodynamically stable. For mixtures such as crystallization cocktails, i.e. the mixture of chemicals designed to drive crystallization, the phase diagram is multi-dimensional. While not true in general, these multi-dimensional macromolecular phase diagrams can often be projected on the temperature-macromolecular concentration plane. This projection changes the temperature scale, however, as the contents of the cocktail such as co-solutes alter the energy scale of macromolecular interactions (while the experimental temperature itself is barely changed). The composition of the final phases can also be complex. A macromolecular crystal, for instance, contains on average more than 50\% of the mother liquor \cite{w15} and may embed co-solutes in fractions that differ from what is left in the crystallization cocktail. 

\begin{figure}[ht]
\centering
\includegraphics[width=\columnwidth]{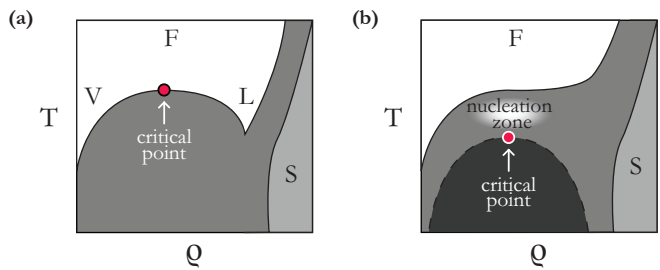}
\centering
\caption{Phase diagram of (a) a simple liquid and (b) a simple protein in the density-temperature plane. Vapor, liquid, solid, and fluid phases are denoted by V, L, S, and F, respectively. For macromolecules, the critical point lies below the solubility line; for simple liquids, this critical point is located at the top of the (hidden) solubility curve. A supersaturated protein solution is typically found to be most likely to crystallize in the nucleation zone -- the region between the solubility line and the critical point.
\label{fig:phase}}
\end{figure}

The arduous task of experimentally determining phase diagrams was carried out for a handful of proteins, including lysozyme, $\gamma$-crystallins, insulin, and myoglobin \cite{w16,w17,w18,w19,w20,w21,w22}. Despite the complexities introduced by the macromolecular structure and the additives in the crystallization cocktail, the phase diagrams of these proteins resemble that of simple liquids \cite{w23}, in the sense that phases analogous to the crystal, liquid, vapor, and supercritical fluid are distinguishable. The phase diagram of simple liquids (shown in Figure~\ref{fig:phase}a), however, exhibits a vapor-liquid coexistence line that terminates at a stable critical point. Below this line (in the dark gray area) the system is unstable, and above the critical temperature, it is a supercritical fluid. One key difference is that for proteins the critical point typically lies below the crystal solubility line, meaning that the protein vapor and liquid phases are metastable. For proteins (Figure~\ref{fig:phase}b), below the critical point, the system has a propensity to aggregate in a disordered, percolating network, i.e., a gel; the gelation probability increasing with distance to the critical point \cite{w24,w25}. Although such an aggregation is metastable, it can be long-lived, which arrests the crystallization process in the timescale of experiments. The area between the solubility line and the critical point is where crystallization is most likely to occur, and has consequently been dubbed the nucleation zone or crystallization gap \cite{w26,w27}.

Macromolecular phase diagrams are informative of the conditions in which crystallization is most likely to occur. However, their experimental determination is a task more challenging and time-consuming than even the most ambitious crystallization screen. Furthermore, theoretically determining phase diagrams to guide crystallization experiments suffers from the fundamental caveat that the macromolecular structure is required, because effective macromolecular interactions cannot be reliably determined otherwise \cite{w63,Fusco:2016}. Therefore, screening remains the most efficient way of probing the space of solution conditions \cite{w28}, even though it can be done only very sparsely. Gaining useful insights about macromolecular crystal assembly requires extracting the key features of the phase diagrams from the limited information provided by experimental screening, and hope to use insights from simpler models to bridge the vast remaining gaps. Hence, although we have a reasonably good understanding of crystallization, the chemical space that needs to be sampled to make use of it is remarkably vast.

\section{Experimental Crystallization Data}
The protein data bank (PDB) is the international archive for all macromolecular structural models and, for crystallography, the X-ray diffraction data used to build these models. For every model in the PDB, a record exists of the crystallization conditions within an unstructured and non-mandatory remark field. Attempts have been made to enhance the usefulness of this record by standardizing chemical definitions and by enabling data queries \cite{w29}. The remarks themselves have also been analyzed revealing that almost 40\% of the reported crystallization conditions can be formed by the optimization of a commercial screen \cite{w30}. Yet, although the PDB remains the largest source of successful crystallization data, over 30\% of its structures contain no crystallization details. When those details are present, they are not systematically recorded, and of course, only the condition used to produce the structural model is captured. Despite these caveats, the information present is clearly useful. For example, Kirkwood et al. \cite{w31} used “cleaned” PDB data to confirm that acidic proteins (pI $<$ 7) tend to crystallize at 1.0 pH unit above their pI and that basic proteins (pI $>$ 7) tend to crystallize 1.5-3.0 pH units above their pI, in agreement with previous studies \cite{w32}. However, the PDB captures no information on other chemical conditions that may have yielded crystals, the range of chemical space sampled, or the other experimental outcomes that result. This is not surprising as the role of the PDB is to capture the structural model and not the complete experimental results to reproduce the result. Similarly, negative results are rarely reported in the literature. Much information that could be useful to understand the crystallization process itself is, however, lost. 

Only a couple of dedicated crystallization condition databases exist. The Biomolecular Crystallization Database (BMCD) includes the subset of the information already captured in the PDB, i.e., about 43,000 entries \cite{w33}, and the Marseille Protein Crystallization Database (MPCD) \cite{w34} combines an older version of the BMCD with the CYCLOP database (no longer available) having a total of about 15,000 entries. A few studies have used these databases to suggest improved crystallization strategies \cite{w35,w36}, but their overall influence has been limited. Over the last decade, however, a number of crystallization facilities have also been gathering datasets that contain the experimental details for each crystallization attempt, for both positive and negative results. This may open the road to types of analysis. Before detailing these efforts, we first review advances made with the datasets that existed prior to this advance.

\section{Analysis of successful crystallization conditions}
The most accessible and immediate source of information about a macromolecule is whether a model of its structure has been obtained from crystal or not up to this point. Because of the intensity of the decades-long efforts at obtaining such models, that information at least reflects in part the experimental difficulty or facility of crystallizing a target.

\subsection{Prediction of crystallization propensity}
Crystallization targets are typically chosen based on two criteria: (i) their biological interest, and (ii) their likeliness to crystallize. To evaluate the latter, a number of properties, such as the number of homologs that have been successfully crystallized, chain length, the frequency of charged residues, and the number of predicted transmembrane helices have long been taken into account \cite{w37}. More recently, a number of crystallization propensity prediction tools have also been developed using structural information from the PDB and information about the primary sequences that yield crystal structures or not from databases such as TargetDB \cite{w38} and PepcDB (now merged as TargetTrack) as well as UniRef50 \cite{w39}. These tools infer from the primary sequence alone the probability with which a macromolecule can be crystallized, which helps select targets more systematically.

SECRET \cite{w40} was one of the first such tools developed, and was shortly followed by others, including OB-Score \cite{w41}, CRYSTALP \cite{Chen:2007}, and ParCrys \cite{w42}, but some of these tools (SECRET and CRYSTALP) have since been rendered obsolete by their low residue limit. In general, the success of these predictors depends strongly on the dataset with which they were trained. Newer training sets are thought to yield better predictors in part because advances in experimental techniques render crystallizable some sequences that used to be deemed uncrystallizable. Testing older prediction software with newer data thus often reveals their lower predictive power. Some of these tools also check features, such as the number of homologs of a given sequence in the PDB, which change with time. In addition, OB-Score, CRYSTALP2 \cite{w43}, ParCrys, SVMCRYS \cite{w44}, MetaPPCP \cite{w45}, and CRYSpred \cite{w46} all use similar training datasets that, according to Mizianty and Kurgan \cite{w37}, are outdated as they were prepared under the assumption that the sequences for which the experiments have been stopped are uncrystallizable. 

Quantitatively, the reported accuracy of these tools can be measured by:
\begin{equation}
\mathrm{Accuracy = \frac{TP+TN}{TP+TN+FP+FN}}
\end{equation}
where TP and TN are the number of true positives and negatives, and FP and FN are the number of false positives and negatives. Note that this quantity gives, for equal fractions of positive and negative results, a minimal accuracy of 50\%. The Matthews correlation coefficient (MCC),
\begin{equation}
\scriptsize{\mathrm{MCC = \frac{TP\times TN -FP \times FN}{\sqrt{(TP+FP)\times(TP+FN)\times(TN+FP)\times(TN+FN)}}}}
\end{equation}
is also used, where a MCC=0 implies that the prediction method does not do better than randomly picking outcomes, while MCC=1 implies a complete predictive power.

In what follows, we detail some of the tools available:

\begin{itemize}
\item XtalPred calculates and predicts several protein properties from the primary sequence, including the average hydrophobicity, pI, the frequency of cysteines, methionines, tryptophans, tyrosines, and phenylalanines, disordered regions, and transmembrane helices. The chain is then assigned a score between one and five by comparing these properties to the distribution of properties in TargetDB. A score of five corresponds to very difficult to crystallize and one means “optimal” for crystallization \cite{w47}. This prediction system was later expanded by using random forest classifiers and by including surface ruggedness, surface entropy and the surface amino acid composition to the list of protein properties. It was also renamed XtalPred-RF \cite{w48}. Note that the use of random forest classifiers was first prototyped in RFCRYS \cite{w49}. This method was selected because it was found to perform better compared to support vector machine (SVM) and artificial neural network methods. Both XtalPred and XtalPred-RF have accuracies around 70\%, but the latter has a higher MCC of 0.47.

\item PPCpred \cite{w37} and PredPPCrys \cite{w50} use a SVM classifier of exposed and buried amino acid compositions, chain disorder, the proximity of certain groups of amino acids and physicochemical properties of individual and collocated amino acids. This large set of features ($\sim$800 for PPCpred and $\sim$3000 for PredPPCrys) is then reduced to obtain a best-performing set. In addition to predicting crystallization probabilities given a sequence, both PPCpred and PredPPCrys also predict the success of other experimental steps, such as purification and the production of well-diffracting crystals. PredPPCrys uses an enhanced feature selection process, a two-level SVM classification system, and an updated dataset. The PredPPCrys server can predict a protein’s propensity of yielding well-diffracting crystals with 76\% accuracy and an MCC of 0.43, while PPCpred, when analyzed with the same datasets, yields an accuracy of 64\% and an MCC of 0.25.

\item SVMCRYS is an SVM-based method \cite{w44} that first used a set of 116 protein features, including amino acid frequencies (individually, in clusters and by amino acid type) and physicochemical properties, such as solvent accessible surface area, hydrophobicity and side chain hydrophobicity. A reduced set of significant features was then identified. As stated above, this method uses an earlier training dataset, and hence its prediction accuracy drops when tested with newer datasets \cite{w50}. The reported accuracy was 87\%, with MCC=0.74. It should also be noted that when the same test set is used to assess XtalPred, the accuracy (78\%) and MCC (0.58) were higher, while these values for ParCrys and OB-Score did not differ significantly from what was originally reported.

\item PPCinter \cite{w51} is another SVM-based predictor that predicts separately the outcome of the same experimental steps as PPCpred. This method uniquely takes into account inter-protein properties, such as the amino acid composition in protein regions predicted to be involved in protein-protein interactions. The reported accuracy (80\%) and MCC (0.55) are about 3\% and 0.03 higher, respectively, than PPCpred.

\item $P_{XS}$ assigns a crystallization probability to proteins by incorporating features that were found to be independent factors impacting crystallizability informed by logistic regression studies, namely the fraction of disordered residues, the average side chain entropies of exposed residues, the fraction of buried glycines, and the fraction of phenylalanine \cite{w52}. By contrast to the other prediction servers mentioned here, this system uses both the structure of crystallized proteins and their experimental propensity of crystallization success. However, this tool, as well as MCSG Z-score \cite{w53}, is limited by its use data from only one structural genomics center for training, and may thus perform poorly in predicting crystallizability of proteins that differ substantially from those in this relatively small training set. Logistic regression also cannot capture non-monotonic trends \cite{w54}. 

\item XANNpred is an artificial neural network trained on a dataset from both PDB and sequence data \cite{w55}. In total 428 protein properties were considered, including pI, average hydrophobicity and the frequency of neighboring pairs of amino acids. The accuracy for this method is between 75\% and 81\% and MCC ranges from 0.50 to 0.63.

\item SCMCRYS is an ensemble method that creates “scoring cards” based on the identity of amino acid pairs separated by 0 to 9 residues as set of features \cite{w56}. The amino acid pairs are associated with a score derived from the frequencies of the pair in crystallizable and non-crystallizable proteins in the training set. These scores are then optimized to maximize the predictive power, and used to predict the crystallizability of the given protein sequence. This method uses the newer training datasets of PPCpred. Its accuracy is found to be about 76\% and MCC about 0.44. Its authors note, however, that the SVM-based PPCpred has a slightly better performance, because it includes more features and SVM can obtain more precise decision boundaries. One of the main purposes of this method is to create more easily interpretable results than those from machine learning.
\end{itemize}

In order to compare performances of the available predictors on an equal footing, it is crucial that the same dataset (consisting of a wide variety of proteins) is used to calculate accuracy, MCC, and other such measures. Without such a comprehensive study, singling out a best method for prediction is difficult. Nevertheless, based on their relatively new training datasets, the wide range of protein features considered, and the appropriateness of the statistical methods chosen, tools such as PredPPCrys, PPCinter, and XtalPred-RF are expected to give better guidance in selecting crystallization targets.

One should, however, keep in mind that most of these methods use structural properties predicted from their sequence alone, a process that is prone to error and thus fundamentally limits the accuracy of the predictors (and why crystallization is being attempted). This limitation is reflected in the relatively low MCC they produce. Although they provide guidance, these tools are far from being complete solutions to predicting outcomes of crystallization trials. Marked improvements require a better understanding of the microscopic properties that promote crystallization and of the crystallization mechanisms themselves, in addition to the use of continually updating training datasets.

\subsection{Statistical analysis of databases and side chain entropy reduction}

The structural models in the PDB have also been surveyed to identify macromolecular features that affect crystallization and thus more specifically guide mutagenesis strategies. The most developed of these approaches is the side chain entropy reduction (SER) method. The method is motivated by the observation that immobilizing side chains at crystal contacts between macromolecules has an entropic cost than can disfavor their formation. This idea is also statistically supported by an analysis of the correlation between protein structures and crystallizability \cite{w54,w57}. SER thus advocates mutating residues with high side chain entropies (SCE), such as lysine, to lower SCE residues, such as alanine. The approach was first successfully applied to Rho GDP-dissociation inhibitor, chosen for its high lysine and glutamate content and low crystallization propensity \cite{w58}. 

Although high SCE residues are predominantly found at the surface of macromolecules, choosing among possible mutagenesis sites from the input amino acid sequence alone is generally a non-trivial task. A more sophisticated algorithm for SER was thus developed, and implemented on the SERp server. It suggests residues for mutation based on: (i) predictions of the secondary structure and the surface SCE, and (ii) sequence homology to identify residues poorly conserved by evolution, and thus likely less functionally or structurally relevant \cite{w59}. Residues are individually scored based on these two criteria and, adjacent, high-scored residues are considered potential sites for mutagenesis. Because removing too many high SCE residues may reduce the solubility of the macromolecule \footnote{In addition to being potentially deleterious to protein solubility, such mutations could also reduce the very structural stability of the macromolecule. Nevertheless, it has been shown that stability does not correlate with its crystallizability, unless the protein is either unfolded (harder) or “hyperstable” (easier) \cite{w52}},
the algorithm also minimizes the number of proposed mutations for a given SER target.  In the end SERp identifies mutations to lysine, glutamate, and glutamine residues. Although successful mutagenesis cases have also mutated asparagine, SERp does not allow it because that residue can also engage in favorable protein-protein interactions. For a number of test cases, the SER strategy has been shown to lead to the crystallization of otherwise recalcitrant macromolecules, and to improve the crystal quality of others \cite{w60}. Thus, protein engineering, of which SER mutagenesis is an example, is one of the means with which the experimental outcome can be improved \cite{w58,w61,w62}.

It should be noted that, although SER help attack the crystallization problem, the physicochemistry of macromolecular crystallization is more intricate than the simple label suggests. For instance, mutating lysine to arginine is likely an alternative for the lysine to alanine mutation – despite them having nearly the same SCE – because favorable electrostatic interactions that enthalpically drive the formation of crystal contacts (with a nearby aspartate, for instance) could otherwise be lost \cite{w54,w60,w63}. Experimental studies indeed show that the interaction enthalpy of the aspartate-arginine pair is ΔH=-6.3 kJ/mol, but only ΔH =1.3kJ/mol for the aspartate-lysine pair \cite{w64}, which is partly due to the aspartate-arginine pair’s ability to form two hydrogen bonds, compared to a single hydrogen bond between aspartate and lysine \cite{w65}. In addition, hydrogen bonds, whose energy can range from $\sim$2 kJ/mol to $\sim$20 kJ/mol  \cite{w66}, can more than compensate the high SCE cost of binding with the backbone. These strong enthalpic contributions thus make it less surprising that some crystal contacts form despite their high SCE cost.

It should also be noted that while changing biochemical conditions for crystallization is straightforward, altering the macromolecule itself—as SER suggests—requires that it can be expressed, and therefore needs additional effort. Although it may have a dramatic impact on outcome, it is certainly not the easiest experimental route to follow.

\section{Analysis of complete crystallization screening data -- success and failure}
Although sparse and imperfect, crystallization screening, either through commercial kits or individual laboratory efforts, remains the most efficient method of probing the phase diagram \cite{w28}. However, little to no information (other than overall success or failure) has been extracted from this screening process in general. In part, this is within the very nature of the scientific enterprise as most studies aim not to understand the fundamentals of crystallization, but rather to obtain a structural model of the macromolecule or complex studied. Compounded to this cross-purpose, only the final crystallization condition is usually reported - any initial screening results are masked by the process of optimization where an iterative process refines the initial promising screening result.  Access to a data or a database with complete and detailed experimental information would allow a more complete study of the effect of chemical conditions on crystallization probability and outcome in relation with macromolecular properties. Given the limited but important information that has been gained from a binary crystal or no crystal analysis of limited crystallization data, the results of such analyses could be instructive about how to optimize screen conditions and a lot more. We detail below two types of insights that have been teased out of this analysis as well as means to push this general idea forward.

\subsection{New biological knowledge}
We know that chemically related conditions are likely to result in similar outcomes and, conversely, chemically distinct conditions are likely to produce different outcomes. This observation is fundamental to the design of different crystallization screens for different classes of targets. This knowledge can be used to extract information from crystallization screening of a single macromolecular target if the chemical conditions can be mathematically described and compared.

\begin{figure}[ht]
\centering
\includegraphics[width=\columnwidth]{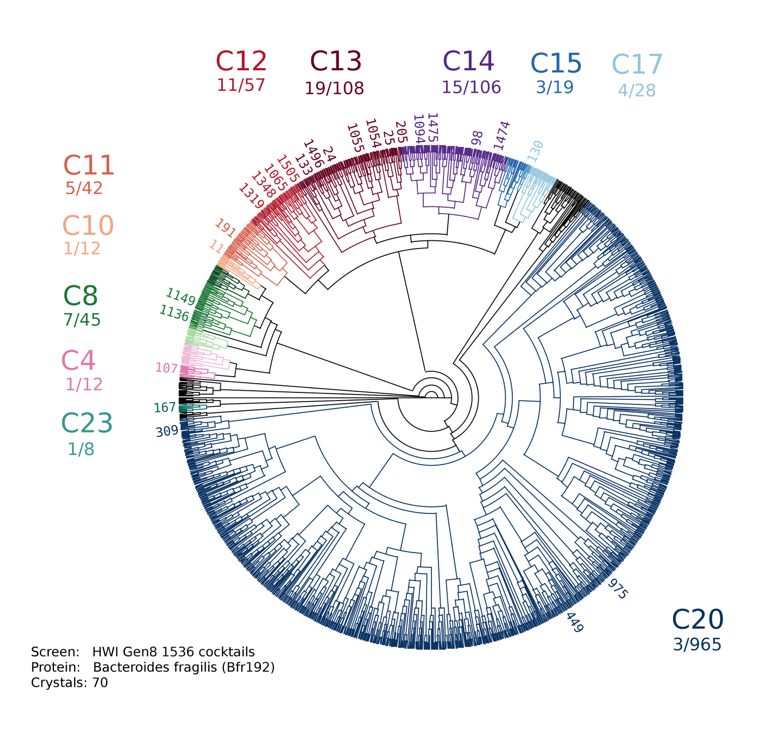}
\centering
\caption{Regions of crystallization space where hits for BfR192 were found. Out of the 28 clusters, 11 were identified containing at least one result determined visually to be crystalline. Adapted from Bruno et al. \cite{w68}. \label{fig:fig2}}
\end{figure}

Newman et al. \cite{w67} pioneered a similarity metric, termed the C6 metric, that assigns a quantitative value to the similarity between two or more crystallization cocktails, and allows those that are chemically similar (through obvious, or less-apparent relationships) to be distinguished from those that are chemically distinct. Using similar methods, related cocktails can be displayed with hierarchical clustering and experimental outcomes overlaid on the cluster \cite{w68}. Applying this approach to an exopolyphosphatase-related protein from Bacteroides fragilis, (BfR192) showed that biological information is contained within the collection of crystallization screening outcomes. A fact that may not be apparent from considering a single outcome alone – successful or not. As an example of this effect, BfR192, is a 343 residue protein with a molecular weight of 39.77 kDa that was screened in our laboratory using 1,536 different cocktails. Figure \ref{fig:fig2} overlays the location of specific cocktails that produced crystals on the clustering of 1,536 different biochemical conditions. Out of the 28 clusters, 11 produced at least one crystal hit.  

Analyzing the outcomes, 70 conditions out of the full 1,536 produced initial crystallization hits. The highest frequency of hits was in cluster 13 followed by cluster 14, and then by cluster 12. In terms of the percentage of hits, cluster 12 ranks as the first with 19\% followed by cluster 13 with 18\% yielding hits. (The total number of cocktails in each cluster is variable due to the design of the screen incorporating commercial screens which operate on differing principles, e.g. grid screening, identifying particular chemical species, the use of multiple small molecules, cryogenic compatibility or incomplete factorial sampling of chemical space as used by the non-commercial condition sampling). Linking the clusters back to their chemistry, cluster 13 proved interesting in that sodium was present in 73\% of the conditions as opposed to 47\% for the 1536-condition-screen overall, potassium was present in 72\% of the conditions versus 24\% overall and finally phosphate was present in 100\% of the conditions versus 16\% overall. 

\begin{figure}[ht]
\centering
\includegraphics[width=\columnwidth]{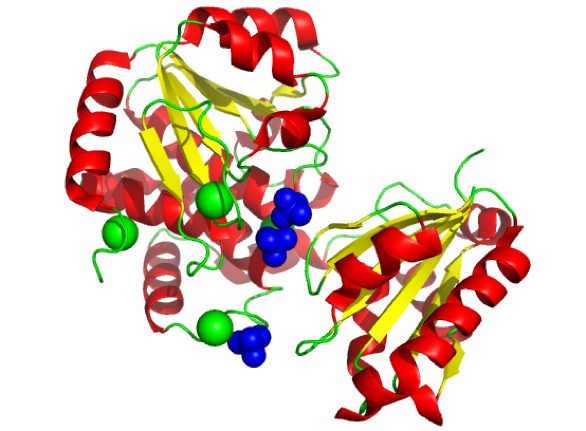}
\centering
\caption{Structure of the BfR192 exopolyphosphatase-related protein (PDB ID 4PY9) showing the two domains and highlighting the cleft containing the sodium, potassium and four phosphate ions \cite{w68}. \label{fig:fig3}}
\end{figure}

The initial structure was deposited in the PDB with undefined electron density. Based on the clustering analysis a sodium ion was placed in the structure, Figure \ref{fig:fig3}, which improved the electron density as did the inclusion of a potassium ion and of four phosphates. The R and Rfree reduced from 22.3\% and 25.9\% to 20.7\% and 24.3\%, respectively. These ions also proved to be biologically relevant. The location of the phosphate-binding pocket suggests that the phosphoryl moieties of polyP might anchor in this pocket. The putative active site has features that are consistent with active sites of other phosphatases involved in binding the phosphoryl moieties of nucleotide triphosphates \cite{w69}. A possible role of the active site phosphate is to ensure a proper substrate orientation and polarization of the phosphoryl P-O bond to increase the susceptibility of the P atom to nucleophilic attack. The space around the phosphate ions suggests that the cleft can bind longer polyP substrates. An interesting observation was the lack of significant crystal formation in the PEG-based conditions (cluster C20), demonstrating one of the two classes of macromolecules revealed in the next section.

In short, the analysis of crystallization results from the initial screen aided by a mathematical description of chemical cocktail similarity identified the presence of ions and a putative biological function.  Even for a single target and only considering crystal or no-crystal outcomes, the analysis of the crystallization results revealed new biological information.

\subsection{Analysis of crystallization propensity}
From the description of the prediction schemes detailed in Section 4, it is natural to wonder what cocktail properties lead to higher crystallizability. In order to productively conduct such a statistical analysis, it is crucial that the dataset assayed systematically contains both positive hits and negative results of crystallization experiments, as well as details of the experimental conditions. 

In this context, the Northeast Structural Genomics consortium (NESG) database, which contains both the crystallization outcome and the full description of chemical conditions for 1,536 cocktails, is an ideal model. From an analysis of a subset of its content, one could statistically reconstruct the experimental crystallization propensity, $\xi$, – the fraction of hits – based on structural features of the macromolecules alone \cite{w54}. Such a reconstruction revealed two mechanisms for crystallization: (i) driven by SCE and, (ii) reliant on specific electrostatic interactions. This nuance was also suggested by Price et al., in a study of a subset of the NESG database \cite{w52}, although the statistical signature was not as clear. Interestingly, this further dataset enabled studying the response in $\xi$ upon varying chemical conditions, which revealed that proteins identified to crystallize with the second mechanism have a propensity to crystalize with changes in salt concentration rather than molecular crowding caused by PEG, for example. 

\section{Technical Considerations for Computational Crystallization}
In order to expand the scope and capability of the above analysis, we discuss here four technical considerations.

\subsection{Systematicity and completeness of datasets}
A key prerequisite of the dataset for the study discussed in Section 5.2 was that $\xi$ was obtained from experiments that span the same chemical space. Comparing results for different screens would be less reliable, and changing the number of screens would bias the “resolution” of $\xi$, especially when it is small. The details of the chemical composition are also essential. While some steps have been made to account for this in other studies with similarity metrics employed \cite{w54,w67}, more detailed knowledge of the chemical processes has to be captured. For instance, the counter-ion identity can be important to consider as different species can have different salting-out abilities. Proper bookkeeping of chemical conditions and individual and combined properties also allows efficient parsing of the data. Because increasing the number and diversity of proteins in the dataset should help reveal subtler trends, this effort could have a significant payoff.

\subsection{Crystallization technique}
For the studies described in Section 5, the initial method of crystallization was microbatch-under-oil. With the exception of concentration changes due to dehydration through the plate, the microbatch method is a static one, there is no dynamic change of component concentration as occurs in vapor diffusion, the predominant crystallization method.  Although no systematic study of the matter has been conducted, one may refer to the physical basis of crystallization for insights. Under-oil experiments keep the solution volume constant, while vapor diffusion gradually supersaturates the protein solution as the droplet volume shrinks. Yet the typical increase in concentration of salt and other co-solutes change the pH and the chemical potential of the co-solutes relatively little, because both depend logarithmically on solute concentration. The protein concentration, however, may change sufficiently for the system to cross the phase boundary into the crystallization regime and to significantly lower the barrier to nucleation (Figure \ref{fig:phase}). Hence, although the volume change and the consequent change in the chemical properties of the solution can likely be ignored in the analysis as a first approximation, the change in macromolecular concentration likely cannot.

\subsection{Optimization of Crystallization Conditions Based on Screening}
Commercial kits developed over many years of experience have made the initial crystallization screening process relatively easy for most laboratories \cite{w7}. The development of such screens involves taking an initial promising condition and changing the chemistry to optimize that condition. This, however, has two shortcomings: (i) optimization occurs around common fixed starting points, and (ii) focus may be on the single best initial result, causing separate but distinct clusters, as revealed in the example from Section 5.2, to be overlooked. Condition optimization is thus an important aspect that requires attention in analyzing crystallization outcome as it can impart an unintentional yet non-fictitious bias to the results.

\subsection{Characterizing outcome}
To best utilize the theoretical knowledge of crystallization, the experimental outcome has to be classified into distinct regions of the protein phase diagram. The phase diagrams obtained for model proteins (Section \ref{sec:crys}) relied on very defined crystallization screens with few variables. In reality, screening a structurally uncharacterized macromolecule probes multiple variables. Given the number of experiments required to cover that chemical space, considerable effort is required to identify potential crystal conditions, let alone other outcomes. When these outcomes are noted for a single macromolecule, they can be used to construct a vector for crystallization within the surveyed chemical space \cite{w70}. However, human classification becomes an impediment when a multitude of macromolecules are studied.

In a detailed study led by one of us, around 150,000 images of the crystallization plates from 96 distinct macromolecules were visually classified by a team of eight viewers such that each image was surveyed by a minimum of three viewers. A control set unknown to the classifiers was included at the beginning, middle and end of the classification effort. Each image was visually classified into seven categories: clear, phase separation, precipitate, skin, crystals, junk, and unsure. Multiple classifications were allowed for all but “clear”. This not only created a training set for automated image classification attempts, but also revealed the reliability of humans in the process. Using a deliberately strict definition of common classifications (classifications had to be exactly equal), the average agreement for classifications of the control set was 78\%. When comparing the whole data set, approximately 48\% of the images were unanimously classified by three separate viewers. Again a deliberately strict interpretation of agreement was used where multiple classifications, e.g. skin and precipitate, were regarded different from either skin, precipitate or the combination and a third component. This work establishes a baseline for automated classification success \cite{w4}. Using this training set and extensive feature extraction approach, automated classification can successfully recognize 80\% of crystal containing images, 94\% of clear images and 94\% of precipitate. Although this approach is promising, the runtime is on the order of seconds per image using high-end graphical processing units \cite{w71}, which limits its large-scale deployment.

Many other attempts have been made in automating crystallization screening classification. To date none are routinely available. A limiting factor in making use of the complete information produced by crystallization screening is the efficient characterization of outcome.

\section{Outlook for the future}
A fairly simple analysis of crystallization screening outcome from multiple experiments on a single macromolecule shows that useful biological information can be obtained. Not discussed here, but also a clear potential, is the ability to distinguish between different states of a macromolecule that may be present as a function of common crystallization conditions. A more global analysis on multiple macromolecules reveals two broad categories that correspond to salt-based crystallization approaches or other classes of precipitants. Membrane-bound macromolecules can be predicted and distinct crystallization approaches used in those cases. These observations, coupled with the success of commercial screens targeting specific macromolecular groups, e.g. nucleic acids, suggest that crystallographic screening results contain both real and useful information. 

The theory of macromolecular crystallization is increasingly well developed. Phase diagrams are fairly well understood but the chemical space that needs to be sampled for a biological macromolecule remains humongous. With a more comprehensive description of chemical components and their interactions, experimental outcome classified on the basis of phase diagram regions, and accurate information on the macromolecular sample including those that fail, successful predictive techniques for both screening and optimization should be within reach. Admittedly, these requirements are not trivial. Although they are not necessary for producing an individual structural model, they are essential to advance the field of crystallization, so that many more structural models can be provided. 

With only one in ten soluble macromolecules on average yielding a structural model today, advances in crystallization success will likely have to be more dramatic than deriving appropriate chemistry for optimizing conditions. Mutation of the sample to improve the crystallization properties appears to be the next step following biochemical or biophysical (e.g. temperature) approaches. A full computational analysis of initial crystallization screening would be more robust in determining the most recalcitrant samples and prioritizing the molecular biology approach.

Computational crystallization approaches are useful in many cases. Achieving their full potential requires significant increase in the amount and quality of available data. To make this happen our high-throughput laboratory and partners in similar laboratories around the globe, which have privileged access and responsibility to collect and curate these datasets, aim to keep on pioneering the analysis and the best practices. Yet, this work can be a challenge to motivate in a conservative funding context. The potential for computational crystallization is thus here but remains to be fully realized.

\begin{acknowledgements}
We acknowledge support from National Science Foundation Grant No. NSF DMR-1055586 and the National Institutes of Health, R01 GM088396.
\end{acknowledgements}
\bibliography{final}
\bibliographystyle{rsc}

\end{document}